\documentstyle[aps,prl,psfig]{revtex}
\begin{document}
\wideabs{
\title{The normal state Fermi surface of pristine and Pb-doped Bi$_2$Sr$_2$CaCu$_2$O$_{8+\delta}$ from ARPES measurements and its photon energy independence.}
\author{S. Legner$^1$, S. V. Borisenko$^1$, C. D\"urr$^1$, T. Pichler$^{1,2}$,  M. Knupfer$^1$, M. S. Golden$^1$,  J. Fink$^1$}
\address{$^1$ Institute for Solid State Research, IFW Dresden, P.O. Box 270016, D-01171 Dresden, Germany}
\address{$^2$ Institut f\"ur Materialphysik, Universit\"at Wien, Strudlhofgasse 4, A-1090 Wien, Austria }
\author{G. Yang, S. Abell}
\address{School of Metallurgy and Materials, The University of Birmingham, Birmingham, B15 2TT, United Kingdom}
\author{H. Berger}
\address{Institut de Physique Appliqu\'ee, Ecole Politechnique F\'ederale de Lausanne, CH-1015 Lausanne, Switzerland}
\author{R. M\"uller, C. Janowitz}
\address{Institut f\"ur Physik der Humboldt Universit\"at zu Berlin, Invalidenstr. 110, D-10115 Berlin, Germany}
\author{G. Reichardt}
\address{BESSY GmbH, Albert-Einstein-Str. 15, D-12489 Berlin, Germany}
\date{\today}
\maketitle
\begin{abstract}

We address the question as to whether the topology of the normal state Fermi surface of Bi$_2$Sr$_2$CaCu$_2$O$_{8+\delta}$ - as seen in angle resolved photoemission - depends on the photon energy used to measure it. High resolution photoemission spectra and Fermi surface maps from pristine and Pb-doped Bi$_2$Sr$_2$CaCu$_2$O$_{8+\delta}$ are presented, recorded using both polarised and unpolarised radiation of differing energies. The data show clearly that no main band crosses the Fermi surface along the $\Gamma$$\overline M$Z direction in reciprocal space, even for a photon energy of 32 eV, thus ruling out the existence of a $\Gamma$-centred, electron-like Fermi surface in this archetypal high T$_C$ superconductor. The true topology of the normal state Fermi surface remains that of hole-like barrels centred at the X,Y points of the Brillouin zone. 

\end{abstract}

}

There is currently an ongoing and lively discussion as regards the true topology and character of the normal state Fermi surfaces (FS) of the high temperature superconductors (HTSC) in general, and of Bi2212 in particular. 
The 'traditional' picture seen in angle-resolved photoemission spectroscopy (ARPES) is that of three different features with different origins:
the main FS centred around the X(Y) points \cite{XYM} of the Brillouin zone (BZ), as predicted by band structure calculations \cite{bandstructure};
the so-called shadow FS due to antiferromagnetic spin correlations \cite{Aebi};
and extrinsic features (diffraction replicas (DRs)) which result from a diffraction of the outgoing photoelectrons as they pass through the structurally modulated Bi-O layers \cite{Ding}.

Recently, ARPES data recorded with photon energies of 32 - 33 eV seemed to show a different picture and have been interpreted in terms of an 
electron-like FS centred around the $\Gamma$ point \onlinecite{Chuang,Feng,Zakharov}.
It has even been suggested that the ARPES-derived FS depends on the photon energy used in the experiment \cite{Feng}.
This, of course, would constitute a revolution in our thinking about the normal state FS of the HTSC and thus it is of utmost importance that this question be addressed quickly and clearly by a number of independent groups.
In this contribution, we present ARPES investigations of Bi2212, with the aim of clearing up the controversy regarding the apparant photon energy dependence of normal state FS topology as seen by photoemission.

We present a combination of energy distribution curves (EDCs) measured using polarized synchrotron radiation with angle-scanned photoemission data using unpolarised radiation at various photon energies and demonstrate that, as physical intuition dictates, the main FS of the Bi2212 materials is {\it independent} of the photon energy used to measure it in an ARPES experiment.

The synchrotron-based data were recorded using the U2-FSGM beamline at the BESSY I facility, with a sample temperature of 100K, an overall energy resolution of 70 meV and an angular resolution of $\pm$1$^\circ$, which gives $\Delta${\bf k} $\leq$ 0.094 \AA$^{-1}$ (i.e. 8.1 $\%$ of $\Gamma$X) in the case of 32 eV radiation.
In all cases the crystals were aligned such that the high symmetry direction being scanned was parallel to the electric field vector of the incoming synchrotron radiation. For the $\Gamma$Y scans the analyser was then swung downwards out of the plane spanned by the surface normal and the 
E-vector, whilst for the $\Gamma$$\overline M$Z scans the energy analyser remained in the aforementioned plane.
The angle-scanned ARPES experiments were performed at 300 K or 120 K using monochromated, unpolarised He I and He II radiation and a SCIENTA SES200 analyser enabling simultaneous analysis of both the E and {\bf k}-distribution of the photoelectrons. The overall energy resolution 
was set to 30 meV and the angular resolution to $\pm$0.38 $^\circ$, which gives $\Delta${\bf k} $\leq$ 0.028 \AA$^{-1}$ (i.e. 2.4 $\%$ of $\Gamma$X) in 
the case of He I radiation.
High quality single crystals of pristine \cite{Bi2212crystals} and Pb-doped Bi2212, the latter grown from the flux in the standard manner, were cleaved 
in-situ to give mirror-like surfaces \cite{Tc}. 

Returning to the current ARPES controversy - certain points are universally accepted.
Firstly, there is a consensus that the 'traditional' FS picture is correct for ARPES data recorded
with low photon energies (h$\nu$ $\leq$22eV) \cite{Sergey,Feng}
Secondly, with respect to the high symmetry directions in {\bf k}-space, the main FS crossing along the $\Gamma$X direction is also
generally accepted to be valid for all photon energies used to date.
Thus, it is in fact the exact situation around the $\overline M$ point which is central to the debate, as it is in this region of {\bf k}-space where the 'closing' of the main FS arcs to give a $\Gamma$-centred (electron-like) FS has been proposed \cite{Chuang,Feng}. 

Consequently, in order to investigate the validity of the 'new' FS topology in detail, as well as to address the question as to whether the final states (17-20 eV above E$_F$) accessed with lower photon energies are sufficiently high to guarantee their free-electron-like character, we have measured EDCs of Bi2212 using synchrotron radiation of different energies along the $\Gamma$$\overline M$Z line in {\bf k}-space.
The data are shown in Fig. \ref{energy_plot}.
For h$\nu$=32eV, the $\Gamma$$\overline M$ data are very similar to those reported in Ref. \onlinecite{Chuang}, having been recorded in the same experimental geometry.
In particular, the reduction of spectral weight around $\overline M$ for h$\nu$=32eV, and to a lesser extent for 40eV photons, could indeed be seen as a sign of a FS crossing, followed by the re-appearance of the band between $\overline M$ and Z.
However, a reduction of the spectral weight of the states related to the extended saddle-point singularity around $\overline M$ for h$\nu$ around 30eV has been predicted to be due to matrix element effects alone in a recent theoretical treatment \cite{Bansil}.
Furthermore, for h$\nu$=50eV, for which no-one would doubt the validity of a free-electron-like final state, the situation resembles to that at lower photon energy and thus we see no indication of a $\Gamma$$\overline M$ FS crossing.

Since it is the h$\nu$=32 and 40 eV data which most significantly deviate from the commonly accepted picture, we devote the rest of the paper to their detailed discussion. 
The claims for a $\Gamma$$\overline M$Z main FS crossing have been based not only on the intensity suppression around $\overline M$ as seen in Fig. 1, but also on an analysis of the {\bf k}-dependence of both the total photoemission intensity (which is related to the momentum distribution n({\bf k})) and of the magnitude of the ARPES intensity at the Fermi level (I(E$_F$)) \cite{Chuang}. 
Fig. 2a shows data for h$\nu$=32eV for both the $\Gamma$Y (panel 1) and $\Gamma$$\overline M$ (panel 2) directions.
We start first with the uncontroversial $\Gamma$Y direction.
The grey-scale image and I$_{int}$ / I(E$_F$) analysis shown in the panels marked 1 contain the 'signature' of a main FS crossing - with a {\it sharp} peak in I(E$_F$) coinciding with a {\it steep} drop in I$_{int}$. 
However, the analogous data for the $\Gamma$$\overline M$ direction (Fig. 2a - panels marked 2) show a different behaviour: both the drop in the total ARPES intensity as well as the peak in I(E$_F$) are considerably broader than their counterparts along $\Gamma$Y.
In particular, the I(E$_F$) peak is more than a factor of three broader than was the case for the $\Gamma$Y main FS crossing. 
The question then arises as to whether this I$_{int}$ / I(E$_F$) characteristic for $\Gamma$$\overline M$ (h$\nu$=32eV) is compatible with a {\it main} FS crossing.
We believe that it is not, and will lay out our arguments for this in the following.

Firstly, assuming for the sake of argument the validity of the $\Gamma$-centred FS, the data for $\Gamma$Y and $\Gamma$$\overline M$ both represent scans crossing the FS at right angles (see the sketch at the top of Fig. 2a). 
Why, then, should the I$_{int}$ and I(E$_F$) analyses for the two directions be so different?

One argument that immediately springs to mind is based upon the fact that the photoemission features along the two directions (directly seen as white features in the I(E,{\bf k}) images of Fig. 2a) have different dispersion relations, thus possibly leading to the anomalous width in both I$_{int}$ and I(E$_F$) for $\Gamma$$\overline M$.
A stringent test of this argument would be to compare the $\Gamma$$\overline M$ data with the I$_{int}$ / I(E$_F$) characteristics of a band, which not only crosses the main FS at right angles, but also displays the {\it same} dispersion relation as that observed along $\Gamma$$\overline M$ for {\bf k}$\leq${\bf k}$_F$.
Ideally speaking, this test should also be carried out for h$\nu$=32eV, but in practice this is hampered by severe difficulties in the location of a true 
right-angular FS crossing, which could not, of course, be along a high symmetry direction.
This last point means that additional complications in the quantification of I$_{int}$ and I(E$_F$) would also result from the strong matrix-element effects implicit in the use of polarised synchrotron radiation (e.g. h$\nu$=32eV).
Furthermore, the DR features which 'decorate' the ARPES data of pure Bi2212 make it harder still to find a suitable main FS crossing to use as a test system. 

Therefore, in order to determine the I$_{int}$ / I(E$_F$) signature of a main (right angular) FS crossing with dispersion equal to that seen along $\Gamma$$\overline M$ for h$\nu$=32eV we turn to data from Pb-doped Bi2212, measured with unpolarised He I radiation (h$\nu$=21.2eV).
This approach has the following advantages: use of unpolarised radiation minimises the differences in datasets recorded with different azimuthal angles and ARPES data from Pb-doped Bi2212 are simpler to interpret due to the absence of DR features \cite{Aebi_PbdopedBSCCO}.
In Fig. 2b we show the comparison between Pb-doped Bi2212 ARPES data for $\Gamma$Y (panel 3) and for a different direction in {\bf k}-space (roughly from 0.4($\Gamma$$\overline M$) towards Y), representing a right-angular crossing of the main FS (panel 4). 
As is evident from Figs. 2a and 2b, the dispersion relations of the bands in panels 2 and 4 are essentially identical - thus we have found a suitable candidate for our test.
This search was only made possible by the use of the full-EDC FS map shown at the top of Fig. 2b.
The lower panels of Fig. 2b show without any doubt that the I$_{int}$ and I(E$_F$) characteristic of a main FS crossing is essentially unaffected by the steepness of the dispersion relation of the band coming up to the FS as {\it both} panels 3 and 4 of Fig. 2b show sharp peaks in I(E$_F$) coupled to a steep drop in I$_{int}$.
This, then, is in favour of our contention that the h$\nu$=32 eV $\Gamma$$\overline M$ data shown in Figs. 1 and 2a do not signal a main FS crossing in the Bi2212-based materials.

A further argument is based upon an analysis of the binding energy position of the leading edge of the ARPES spectra.
In our experience, based upon full-EDC FS maps comprising more than 4000 spectra \cite{Sergey}, the leading edge of the spectra not only approaches E$_F$ as the band disperses up towards the FS, but also moves rapidly away from E$_F$ again once the band has crossed the FS (this is a consequence of the well-known incoherent background present in ARPES data of all HTSC).
Thus, following the leading edge energy as a function of {\bf k}, a main FS crossing exhibits a sharp dip centred at {\bf k}$_F$, as is illustrated in Fig. 3.
Figs. 3a, 3b and 3c show analyses of the leading edge energy for the h$\nu$=32 eV data for $\Gamma$Y (which is shown in panel 1 of Fig. 2a), and the He I data shown in Fig. 2b (panel 3) and Fig. 2b (panel 4), respectively. 
In all cases, the main FS crossing, and thus {\bf k}$_F$, are characterised by the sharp dip or 'V' in the leading edge energy.
This behaviour is to be compared with that for the $\Gamma$$\overline M$ direction (Figs. 3d-f) in which, regardless of the photon energy, no sharp, 
'V'-like structure is seen in the leading edge energy profiles centred around the proposed FS crossing ({\bf k}$_F$ = 0.81 and 1.19 ($\Gamma$$\overline M$)). Thus it is clear that the leading edge data of Fig. 3d (h$\nu$=32eV) should be grouped with the data of Fig. 3e and 3f which characterise flat-band, 
saddle-point behaviour, and {\it not} with the leading edge datasets describing a main FS crossing (Figs. 3a-3c).

Taking the arguments given above, the viewpoint that the observed \cite{Chuang,Feng} FS 'crossings' along the $\Gamma$$\overline M$Z line in Bi2212 result, in fact, from the superposition of extrinsic DR features\cite{Sergey,Fretwell,Mesot} is considerably strengthened. 
In Ref. \onlinecite{Sergey} we argued that multi-order DRs combine to give a high intensity ribbon, visible in FS mapping data running along the 
(0,-$\pi$)-($\pi$,0) line.
The suppression of the spectral weight from the extended saddle-point singularity states predicted for photon energies around 30 eV \cite{Bansil}, would then lead to a 'hollowing-out' of the ribbon - leaving its edges intense enough to appear as a pair of FS crossings either side of the $\overline M$ point.
In order to test this point, and bearing in mind the efficacy of FS maps recorded using unpolarised radiation and based upon real, uninterpolated EDCs \cite{Sergey}, we have carried out such FS mapping experiments on Pb-doped Bi2212, in which the Pb substitution supresses the incommensurate Bi-O modulation \cite{Aebi_PbdopedBSCCO} and thus 'switches off' the DR features in the ARPES spectra.

Fig. 4 shows the FS maps, in which I(E$_F$) for a 20 meV energy window (T=300K) is plotted.
Data recorded using He I radiation (h$\nu$=21.22eV) are shown in Fig. 4a, whereas Figs. 4b and 4c contain smaller maps measured with He II (h$\nu$=40.8eV) radiation which highlight those areas in {\bf k}-space indicated by the dotted lines in Fig. 4a.
In each case, the main hole-like FS centred at the X and Y points is clearly visible (solid white line). 
While these conclusions are under no doubt for the He I data, upon consideration of Fig. 1, it can be seen that a photon energy of 40eV is still in the critical range for which an intensity suppression near $\overline M$ is observed. 
Thus, we point out that Fig. 4c (h$\nu$=40.8eV) shows no indication of a FS crossing at the points 0.81 and 1.19 ($\Gamma$$\overline M$) as suggested in Ref. \onlinecite{Chuang}, nor at {\it any} point along or near to the $\Gamma$$\overline M$Z line.
Therefore, the FS maps presented in Fig. 4, taken together with the detailed analysis of I$_{int}$ / I(E$_F$) (Fig. 2) and of the leading edge energies of the ARPES spectra (Fig. 3) offer very strong additional support to arguments that the alleged FS crossings along the $\Gamma$$\overline M$Z direction in pristine Bi2212 are, in fact, due to DR features \cite{Sergey,Fretwell}. These dominate the ARPES spectra as a result of the matrix element-related supression of the saddle-point emission near $\overline M$ for photon energies around 30eV \cite{Bansil}.

Thus, in summary we can state that such DR-related 'FS crossings' along $\Gamma$$\overline M$Z in Bi2212 do not have any consequences for the true normal state FS topology of the Bi2212-based HTSC, which remains that of hole-like barrels centred at the X,Y points, independent of the photon energy used in the ARPES experiment.

We are grateful to the the BMBF (05 SB8BDA 6), the DFG (Graduiertenkolleg 'Struktur- und Korrelationseffekte in Festk\"orpern' der TU-Dresden) and the SMWK (4-7531.50-040-823-99/6) for financial support, to U. J\"annicke-R\"ossler and K. Nenkov for characterisation of the crystals. T.P. acknowledges an APART fellowship of the Austrian Academy of Sciences.

\begin{figure}
\caption{ARPES of Bi2212 recorded at 100K along the {$\Gamma$}M direction in {\bf k}-space for the photon energies: (a) 32 eV, (b) 40 eV and (c) 50eV. 
For each photon energy, the right panels show the spectra and the left panels (E,{\bf k})-plots in which the photoemission intensity is represented by a (linear) grey scale. For (a) the displayed region in {\bf{k}}-space goes from  0.19 up to 1.42 \AA$^{-1}$, for (b) from  0.13 to 1.36 \AA$^{-1}$ and for (c) from  0.24 to 1.45 \AA$^{-1}$.}
\label{energy_plot}
\end{figure}

\begin{figure}
\caption{
(a) ARPES data from Bi2212 recorded with h$\nu$=32 eV at 100K. (b) ARPES data from Pb-doped Bi2212 recorded with h$\nu$=21.2 eV at T=120K.
The chains of circles on the (a) BZ sketch and (b) FS map indicate the points in {\bf k}-space in which the ARPES data were measured.
In the BZ sketch (a), the conventional and unconventional FS topologies are marked by solid or dotted white lines, respectively.
Below the ARPES images (the linear grey scale indicates the intensity) is shown the analysis of the total photoemission intensity (I$_{int}$: solid squares, related to the momentum distribution n({\bf k})) and the intensity in a 40 meV window centred on E$_F $ (I(E$_F $): open squares). 
All panels have the same momentum scale.
}
\label{nk_int}
\end{figure}

\begin{figure}
\caption{
Relative energy position of the leading edge of the ARPES spectra for: (a)-(c) the data shown in Fig. 2 in panels 1,3 and 4, repectively; (d) the data shown in Fig 2 in panel 2; (e) and (f) the data shown in Fig. 1(b) and 1(c), respectively.
The dashed vertical lines mark the proposed position of {\bf k}$_F$. For (a)-(c) {\bf k}$_F$ is uncontroversial;
for (d)-(f) the positions of {\bf k}$_F$ for the $\Gamma$-centred, electron-like FS proposed in Ref. [5] are indicated.
Note: both the energy and the {\bf k}-scale are identical for (a)-(f).
}
\end{figure}

\begin{figure}
\caption{
FS maps of Bi-O modulation-free Pb-doped Bi2212 recorded using unpolarised radiation at room temperature. The grey scale indicates the photoemission intensity in a 20meV window centred at E$_F$.
(a) h$\nu$=21.22 eV (He I): The upper [lower] dotted areas indicate the portions of {\bf k}-space examined with h$\nu$=40.8 eV (He II) radiation in parts (b)[(c)] of the figure.
The He I and He II FS maps were from consecutive cleavages of the same single crystal. 
Note the complete absence of any sign of a FS crossing along, or near to the $\Gamma$$\overline M$ line in (c).
}
\label{map}
\end{figure}


\begin{references}
\bibitem{Tc} The T$_c$ (${\Delta}$T) of the Pb-doped and the pristine Bi2212 sample were 68 K(2.1 K) and 86.5 K(1.5 K), respectively.
\end{references}
\end{document}